\documentclass[pdftex,twocolumn,epjc3]{svjour3}
\RequirePackage{graphicx}
\RequirePackage{amsmath}
\RequirePackage{amssymb}
\RequirePackage[numbers,sort&compress]{natbib}
\RequirePackage[colorlinks,citecolor=blue,urlcolor=blue,linkcolor=blue]{hyperref}
\journalname{European Physical Journal B}
\begin{document}

\title{Phase boundary near a magnetic percolation transition %\thanksref{t1}
}


\author{Gaurav Khairnar \thanksref{e1,addr1}    \and
        Cameron Lerch \thanksref{e2,addr1,addr2}   \and
        Thomas Vojta \thanksref{e3,addr1}
}

\thankstext{e1}{e-mail: grktmk@mst.edu}
\thankstext{e2}{e-mail: cameron.lerch@yale.edu}
\thankstext{e3}{e-mail: vojtat@mst.edu}

\institute{Department of Physics, Missouri University of Science and Technology, Rolla, Missouri 65409, USA \label{addr1}
           \and
           Department of Mechanical Engineering and Materials Science, Yale University, New Haven, Connecticut 06520, USA \label{addr2}
}

\date{Received: date / Accepted: date}

\maketitle

\begin{abstract}
Motivated by recent experimental observations [Phys. Rev. {\bf 96}, 020407 (2017)]
on hexagonal ferrites, we revisit the phase diagrams of diluted magnets
close to the lattice percolation threshold. We perform large-scale Monte Carlo simulations of XY and Heisenberg models
on both simple cubic lattices and lattices representing the crystal structure of the hexagonal ferrites.
Close to the percolation threshold $p_c$, we find that the magnetic ordering temperature $T_c$ depends on the
dilution $p$  via the power law $T_c \sim |p-p_c|^\phi$ with exponent $\phi=1.09$, in agreement with classical
percolation theory. However, this asymptotic critical region is very narrow, $|p-p_c| \lesssim 0.04$. Outside of it,
the shape of the phase boundary is well described, over a wide range of dilutions, by a nonuniversal power law with
an exponent somewhat below unity. Nonetheless, the percolation scenario does not reproduce the experimentally observed
relation $T_c \sim (x_c -x)^{2/3}$ in PbFe$_{12-x}$Ga$_x$O$_{19}$. We discuss the generality of our findings as
well as implications for the physics of diluted hexagonal ferrites.
\end{abstract}

%%%%%%%%%%%%%%%%%%%%%%%%%%%%%%%%%%%%%%%%%%%%%%%%%%%%%%%%%%%%%%%%%%%%%%%%%%%%%%%%%%%%%%%%%%%%%%%%%%%%%%
\section{Introduction}
\label{sec:intro}
%%%%%%%%%%%%%%%%%%%%%%%%%%%%%%%%%%%%%%%%%%%%%%%%%%%%%%%%%%%%%%%%%%%%%%%%%%%%%%%%%%%%%%%%%%%%%%%%%%%%%%

Disordered many-body systems feature three different types of fluctuations, viz., static random fluctuations
due to the quenched disorder, thermal fluctuations, and quantum fluctuations. Their interplay can greatly affect
the properties of phase transitions, with possible consequences ranging from a simple change of universality
class \cite{GrinsteinLuther76} to exotic infinite-randomness criticality \cite{Fisher92,Fisher95},
classical \cite{Griffiths69} and quantum \cite{ThillHuse95,YoungRieger96} Griffiths singularities, as well as
the destruction of the transition by smearing \cite{Vojta03a,SknepnekVojta04,SchehrRieger06,HoyosVojta08}.
Recent reviews of some of these phenomena can be found in Refs.\ \cite{Vojta06,Vojta10,Vojta19}.
Randomly diluted magnetic materials are a particularly interesting class of systems in which the above
interplay is realized. Here, the disorder fluctuations correspond to the geometric fluctuations
of the underlying lattices which can undergo a geometric percolation transition between a disconnected
phase and a connected (percolating) phase \cite{StaufferAharony_book91}.

Recently, the behavior of diluted magnets close to the percolation transition has reattracted attention
because of the unexpected shape of the phase boundary observed in the diluted hexagonal ferrite (hexaferrite)
PbFe$_{12-x}$Ga$_x$O$_{19}$ \cite{Rowleyetal17}.
Pure PbFe$_{12}$O$_{19}$ orders ferrimagnetically at temperatures below about 720 K \cite{ALWD02}.
The ordering temperature $T_c$ can be suppressed by randomly substituting nonmagnetic Ga ions for Fe ions in
PbFe$_{12-x}$Ga$_x$O$_{19}$. It vanishes when $x$ reaches the critical value $x_c \approx 8.6$.
This value is very close the percolation threshold $x_p=8.846$ of the underlying lattice\footnote{The
lattice in question is the lattice of exchange interactions between the Fe ions.},
suggesting that the transition at $x_c$ is of percolation type \cite{Rowleyetal17}. Remarkably,
the phase boundary follows the power law $T_c (x) = T_c(0) (1-x/x_c)^{\phi}$ with $\phi=2/3$
over the entire $x$-range
from 0 to $x_c$. This disagrees with the prediction from classical percolation theory
\cite{StaufferAharony_book91,Coniglio81} which yields a crossover exponent of $\phi > 1$
for continuous symmetry magnets, at least for dilutions close to $x_c$.

In this paper, we therefore reinvestigate the phase boundary close to the percolation transition of diluted
classical planar and Heisenberg magnets by means of large-scale Monte Carlo simulations.
The purpose of the paper is twofold. First, we wish to test and verify the percolation theory predictions,
focusing not only
on the asymptotic critical behavior but also on the width of the critical region and the preasymptotic
properties. Second, we wish to explore whether the classical percolation scenario can explain the experimental observations
in PbFe$_{12-x}$Ga$_x$O$_{19}$ \cite{Rowleyetal17}.

Our paper is organized as follows. In Sec.\ \ref{sec:model}, we introduce the diluted XY and Heisenberg models
and discuss their qualitative behavior. Section \ref{sec:percolation} summarizes the predictions of percolation
theory. Our Monte Carlo simulation method is described in Sec.\ \ref{sec:simulations}. Sections
\ref{sec:cubic} and \ref{sec:hexaferrite} report our results for model systems on cubic lattices and for
systems defined on the hexagonal ferrite lattice, respectively. We conclude in Sec.\
\ref{sec:conclusions}.

%%%%%%%%%%%%%%%%%%%%%%%%%%%%%%%%%%%%%%%%%%%%%%%%%%%%%%%%%%%%%%%%%%%%%%%%%%%%%%%%%%%%%%%%%%%%%%%%%%%%%%%%%%
\section{The Models}
\label{sec:model}
%%%%%%%%%%%%%%%%%%%%%%%%%%%%%%%%%%%%%%%%%%%%%%%%%%%%%%%%%%%%%%%%%%%%%%%%%%%%%%%%%%%%%%%%%%%%%%%%%%%%%%%%%%%

Consistent with the dual purpose of studying the critical behavior of the phase boundary close to a magnetic
percolation transition and of addressing the experimental observations in diluted hexaferrites \cite{Rowleyetal17},
we consider two models, viz., (i) site-diluted classical XY and Heisenberg models on simple cubic lattices and
(ii) a classical Heisenberg Hamiltonian based on the hexaferrite crystal structure using realistic exchange
interactions. Comparing the results of these different models will also allow us to explore the universality
of the critical behavior.

\subsection{Site-diluted XY and Heisenberg models on cubic lattices}
\label{subssec:model-cubic}

We consider a simple cubic lattice of $N=L^3$ sites. Each site is either occupied by a vacancy or by a
classical spin, i.e., an $n$-component unit vector $\mathbf{S}_i$ ($n=2$ for the XY model and $n=3$ for
the Heisenberg case). The Hamiltonian reads
\begin{equation}
H = -J\sum_{<i,j>}\epsilon_i\epsilon_j\mathbf{S}_i \cdot \mathbf{S}_j~.
\label{eq:H_cubic}
\end{equation}
Here, the sum is over pairs of nearest-neighbor sites, and $J>0$ denotes the ferromagnetic exchange interaction.
(In the following, we set $J$ to unity for the cubic lattice simulations.)
The quenched independent random variables $\epsilon_i$ implement the site dilution.
They take the values 0 (vacancy) with probability $p$ and 1 (occupied site) with probability $1-p$.
We employ periodic boundary conditions. Magnetic long-range order can be characterized
by the order parameter, the total magnetization
\begin{equation}
\mathbf{m} = \frac{1}{N} \sum_i \mathbf{S}_i~.
\label{eq:m}
\end{equation}

The qualitative behavior of this model as a function of temperature $T$ and dilution $p$ is well understood
(see, e.g., Ref.\ \cite{VojtaHoyos08b} for an overview).
For sufficiently small dilution, the system orders magnetically below a critical temperature $T_c(p)$.
The critical temperature decreases continuously with $p$ until it reaches zero at the percolation threshold
$p_c$ of the lattice. For dilutions beyond the percolation threshold, magnetic long-range order is impossible
because the system breaks down into finite noninteracting clusters.
The point $p=p_c$, $T=0$ is a multicritical point at which both the geometric fluctuations of the lattice
and the thermal fluctuations become long-ranged.

\subsection{Hexaferrite Heisenberg model}
\label{subssec:model-hexa}

\begin{figure}
\centerline{\includegraphics[width=6cm]{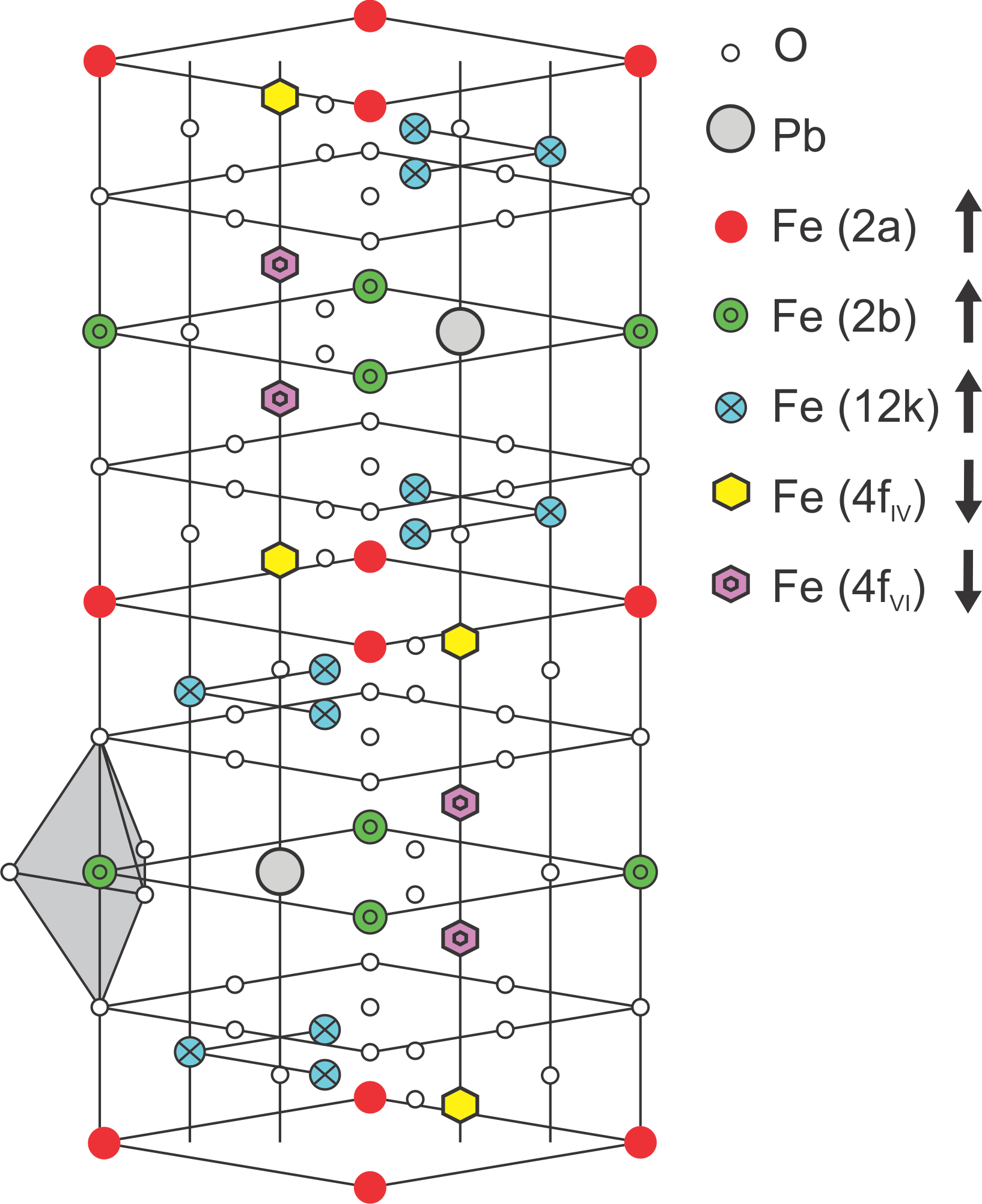}}
\caption{Double unit cell of PbFe$_{12}$O$_{19}$. 24 Fe$^{3+}$ ions are located on five distinct sublattices.}
\label{fig:crystal}
\end{figure}

PbFe$_{12}$O$_{19}$ crystallizes in the magnetoplumbite structure, as illustrated in Fig.\ \ref{fig:crystal}.
A double unit cell contains $24$ Fe$^{3+}$ ions in five distinct sublattices; they are in the spin state $S=5/2$.
Below a temperature of about $720K$, the material orders ferrimagnetically, with $16$ of the Fe spins pointing up
and the remaining $8$ Fe ions pointing down \cite{ALWD02}.
Note that the high critical temperature and the high spin value
suggest that a classical description should provide a good approximation.

In PbFe$_{12-x}$Ga$_x$O$_{19}$, the randomly substituted Ga ions, which replace the Fe ions, act as quenched spinless impurities.
To model this system, we start from the hexaferrite crystal structure and randomly place either a vacancy (with probability $p$)
or a classical Heisenberg spin $\mathbf{S}_i$ (with probability $1-p$) at each Fe site. The dilution $p$ is related
to the number $x$ of Ga ions in the unit cell by $p=x/12$. The Hamiltonian reads
\begin{equation}
H = - \sum_{i,j} J_{ij} \epsilon_i \epsilon_j \mathbf{S}_i \mathbf{S}_j~.
\label{eq:H_hexa}
\end{equation}
The quenched random variables $\epsilon_i$ distinguish vacancies and spins, as before.
The values of the exchange interactions $J_{ij}$ stem from the density functional calculation in Ref.\
\cite{Wuetal16}; they are scaled by a common factor to approximately reproduce the critical temperature
$T_c =720K$ of the undiluted material. In most of our Monte Carlo simulations, we include only the leading (strongest)
interactions which are between the following sublattice pairs: 2a-4f$_{IV}$,  2b-4f$_{VI}$, 12k-4f$_{IV}$, 12k-4f$_{VI}$.
These interactions are non-frustrated and establish the ferrimagnetic order.
We also perform a few test calculations to explore the effects of additional couplings which are significantly
weaker but frustrate the ferrimagnetic order.

The qualitative features of the phase diagram of the model (\ref{eq:H_hexa}) are expected to be similar to those
discussed in the previous section. With increasing dilution $p$, the critical temperature $T_c(p)$ is continuously
suppressed and reaches zero at the site percolation threshold. The value of the percolation threshold of the lattice
spanned by the leading non-frustrated interactions between the Fe ions was determined in Ref.\ \cite{Rowleyetal17}
by means of Monte Carlo simulations. They yielded $p_c=0.7372(5)$, corresponding to $x_c=8.846(6)$ Ga ions per unit cell.
(The numbers in brackets show the error estimate of the last digit.)

%%%%%%%%%%%%%%%%%%%%%%%%%%%%%%%%%%%%%%%%%%%%%%%%%%%%%%%%%%%%%%%%%%%%%%%%%%%%%%%%%%%%%%%%%%%%%%%%%%%%%%%%%%%%%%%%%
\section{Predictions of Percolation Theory}
\label{sec:percolation}
%%%%%%%%%%%%%%%%%%%%%%%%%%%%%%%%%%%%%%%%%%%%%%%%%%%%%%%%%%%%%%%%%%%%%%%%%%%%%%%%%%%%%%%%%%%%%%%%%%%%%%%%%%%%%%%%%
In this section, we briefly summarize the predictions of classical percolation theory for the shape of the phase
boundary $T_c(p)$ close to multicritical point $p=p_c, T=0$ \cite{StaufferAharony_book91,Shender1975,Coniglio81}.
Close to this point, two length scales are at play, the percolation correlation length, $\xi_p$ which characterizes the
size of finite isolated clusters of lattice sites and the magnetic thermal correlation length on the critical infinite
percolating cluster at $p_c$ denoted by $\xi_T$. The percolation correlation length $\xi_p$ diverges as
$\xi_p \sim |p-p_c|^{-\nu_p}$ as the percolation threshold is approached. The magnetic thermal correlation
length behaves as $\xi_T \sim T^{-\nu_T}$ for continuous-symmetry magnets described by the $n$-vector model with $n>1$.

To find the phase boundary, consider the magnetization near the critical point. It fulfills the scaling form,
   \begin{equation}
   m(p-p_c,T) = |p-p_c|^{\beta}~ X\left({\xi_T}/{\xi_p}\right).
   \end{equation}
   For $p<p_c$, the magnetic phase transition occurs at a particular value $x_c$ of the argument of the scaling function $X$. At the magnetic transition, we therefore have $\xi_T = x_c \xi_p$. This yields the power law relation
%    $T^{-\nu_T} \sim |p-p_c|^{-\nu_p}$ or
  \begin{equation}
    T_c(p) \sim |p-p_c|^{\phi}~.
   \end{equation}
   The crossover exponent $\phi$ takes the value $\phi = \nu_p/\nu_T$.
(In contrast, $\xi_T$ diverges exponentially, $\xi_T \sim (e^{-2J/T})^{-\nu_T}$, for Ising magnets, leading to a
logarithmic dependence $T_c(p) \sim \ln^{-1}(1/|p-p_c|)$.)

   Using a renormalization group calculation, Coniglio \cite{Coniglio81} established the relation $\nu_T = 1/\tilde{\zeta}_R$. Here,  $\tilde{\zeta}_R$ characterizes the resistance $R$ of a random resistor network on a critical percolation cluster of linear size $L$ via $R \sim L^{\tilde{\zeta}_R}$.

   The exponent $\tilde{\zeta}_R$ can be related to the well-known conductivity critical exponent $t$ which
   describes how the conductivity $\sigma$ of the resistor network depends on the distance from the percolation
   threshold, $\sigma \sim |p-p_c|^t$. To do so, consider a resistor network on a percolating lattice
   close to $p_c$ but on the percolating side. Its behavior is critical for clusters of size less than $\xi_p$ and
   Ohmic for sizes beyond $\xi_p$. For a $d$-dimensional system of linear size $L \gg \xi_p$,
   we can employ Ohm's law to combine blocks of size $\xi_p$, yielding
   \begin{equation}
   R(L) = R(\xi_p)\left(\frac{L}{\xi_p}\right)\left(\frac{L}{\xi_p}\right)^{-(d-1)}
   \sim ~\xi_p^{\tilde{\zeta_R}} \xi_p^{d-2}  L^{2-d}~.
   \end{equation}
   The conductivity on the percolating side thus behaves as $\sigma \sim \xi_p^{-(d-2+\tilde{\zeta}_R)} \sim |p-p_c|^{\nu_p(d-2+\tilde{\zeta_R})}$.
    Thus, we obtain the hyperscaling relation, $t = (d-2+\tilde{\zeta_R})\nu_p$ or  $\tilde{\zeta_R} = {t}/{\nu_p}-d+2$. Using the numerical estimates $t/\nu_p = 2.28(2)$ and $\nu_p = 0.876(2)$
   \cite{KOZLOV10,Wang13} for three-dimensional systems yields $\tilde{\zeta_R}=1.28(2)$, predicting a
   crossover exponent of   $\phi = \nu_p/\nu_T = \nu_p \tilde{\zeta}_R = 1.12(2)$.
   \footnote{The crossover exponent has also been computed within an expansion in powers of $\epsilon=6-d$ yielding
             $\phi = 1+ \epsilon/42$ to first order in $\epsilon$ \cite{HarrisLubensky84,HarrisAharony89}.
             The resulting value, $\phi = 1.071$, is surprisingly close to the best numerical estimate $\phi=1.12(2)$.}

%%%%%%%%%%%%%%%%%%%%%%%%%%%%%%%%%%%%%%%%%%%%%%%%%%%%%%%%%%%%%%%%%%%%%%%%%%%%%%%%%%%%%%%%%%%%%%%%%%%%%%%%%%%%%%%%%
\section{Numerical Simulations}
\label{sec:simulations}
%%%%%%%%%%%%%%%%%%%%%%%%%%%%%%%%%%%%%%%%%%%%%%%%%%%%%%%%%%%%%%%%%%%%%%%%%%%%%%%%%%%%%%%%%%%%%%%%%%%%%%%%%%%%%%%%%
\subsection{Monte Carlo method}
%%%%%%%%%%%%%%%%%%%%%%%%%%%%%%%%%%%%%%%%%%%%%%%%%%%%%%%%%%%%%%%%%%%%%%%%%%%%%%%%%%%%%%%%%%%%%%%%%%%%%%%%%%%%%%%%%
To find the critical temperature for a given dilution of the system,  we perform large-scale Monte Carlo (MC) simulations.
These simulations employ the Wolff \cite{Wolff89} and Metropolis \cite{Metropolis53} algorithms. Specifically, a full MC
sweep consists of a Wolff sweep followed by a Metropolis sweep. The Wolff algorithm is a cluster-flip algorithm
which is beneficial in reducing critical slowing down of the system near criticality.
The Metropolis algorithm is a single spin-flip algorithm. It is required to achieve equilibration of small
isolated clusters of lattice sites which might form as a result of dilution.

For the cubic lattice calculations, we consider system sizes ranging from $L^3=10^3$ to $L^3=112^3$. We have simulated $4000 - 40000$ independent disorder configurations for each size. For the hexaferrite lattice,
we simulate systems consisting of  $10^3$ to $40^3$ double unit cells (each double unit cell contains
$24$ Fe sites) using $100-300$ independent disorder configurations for each size. All physical quantities of interest,
such as energy, magnetization, correlation length, etc. are averaged over the disorder configurations. Statistical
errors are obtained from the variations of the results between the configurations.

\begin{figure}
\includegraphics[width=\linewidth]{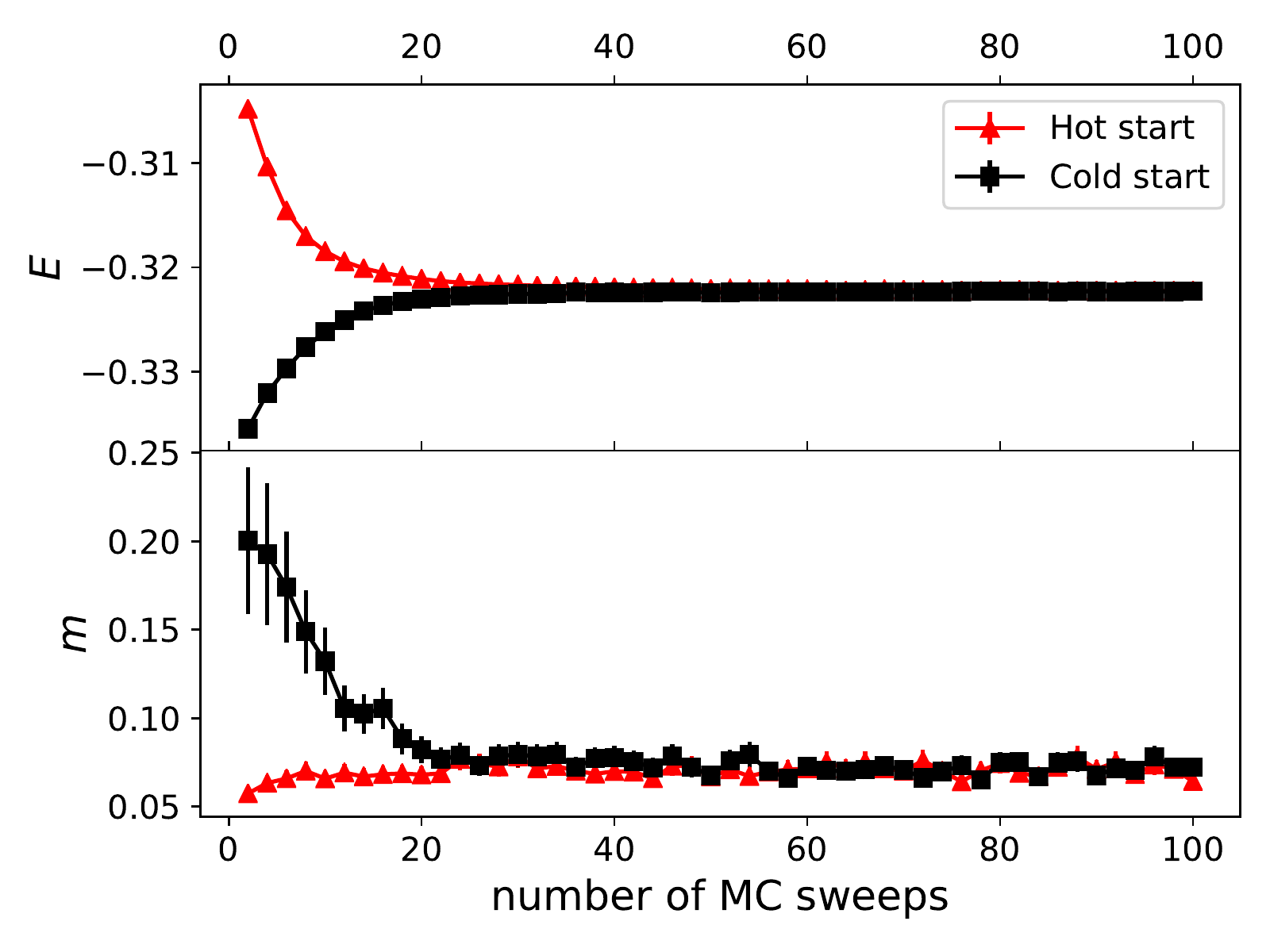}
\caption{Equilibration of the energy per site $E$ and the magnetization $m$ for a cubic lattice XY model of size $L=56$, dilution
$p=0.66$ , and temperature $T=0.156$ averaged over $20$ disorder configurations. The comparison of hot and cold starts shows that the
system equilibrates after roughly $50$ Monte Carlo sweeps despite being close to the multicritical point.}
\label{fig:eqsweeps}
\end{figure}

Measurements of observables must be performed after the system reaches thermal equilibrium.
We determine the number of Monte Carlo sweeps required for the system to equilibrate by comparing the results of
runs with hot starts (for which the spins initially point in random directions) and with cold starts (for which all spins are
initially aligned). An example of such a test for a cubic lattice XY system close to multicritical point is shown in
Fig. \ref{fig:eqsweeps}.
The energy and order parameter attain their respective equilibrium values after roughly $50$ Monte Carlo sweeps.
Similar numerical checks were performed for other parameter values as well as for the cases of Heisenberg spins on cubic and hexaferrite lattices. Based on these
tests, we have chosen $150$ equilibration sweeps (using a hot start) and $500$ measurement sweeps per disorder configuration
for the cubic lattice simulations. For the hexaferrite lattice, we perform $1000$ equilibration sweeps and
$2000$ measurement sweeps (using a hot start). Note that the combination of relatively short Monte Carlo runs and a large
number of disorder configurations leads to an overall reduction of statistical error \cite{BFMM98,VojtaSknepnek06,ZWNHV15}.

%%%%%%%%%%%%%%%%%%%%%%%%%%%%%%%%%%%%%%%%%%%%%%%%%%%%%%%%%%%%%%%%%%%%%%%%%%%%%%%%%%%%%%%%%%%%%%%%%%%%%%%%%%%%%%%%%%%%%
\subsection{Data analysis}
\label{subsec:data_analysis}
%%%%%%%%%%%%%%%%%%%%%%%%%%%%%%%%%%%%%%%%%%%%%%%%%%%%%%%%%%%%%%%%%%%%%%%%%%%%%%%%%%%%%%%%%%%%%%%%%%%%%%%%%%%%%%%%%%%%%
We employ the Binder cumulant \cite{Binder1981} to precisely estimate the critical temperature $T_c$. It is defined as
\begin{equation}
g = \left[1-\frac{\langle|\textbf{m}|^4\rangle}{3\langle|\textbf{m}|^2\rangle^2}\right]_{dis}
\end{equation}
where $\langle...\rangle$ denotes the thermodynamic (Monte Carlo) average and $[...]_{dis}$ denotes the disorder average.
The Binder cumulant $g$ is a dimensionless quantity, it therefore fulfills the finite-size scaling form
\begin{equation}
g(t,L,u) = g(t\lambda^{-1/\nu},L\lambda,u\lambda^\delta)~.
\label{eq:g_FSS}
\end{equation}
Here, $\lambda$ is an arbitrary scale factor, $t = (T-T_c)/T_c$ denotes the reduced temperature, and
$\nu$ is the correlation length exponent of the (magnetic) finite-temperature phase transition.
We have included the irrelevant variable $u$ characterized by the exponent $\delta > 0$ to describe
the corrections from the leading scaling behavior observed in our data. Setting the scale factor
$\lambda=L^{-1}$, we obtain $g(t,L,u) = F(tL^{1/\nu},uL^{-\delta})$ where $F$ is a dimensionless scaling function.
Expanding $F$ in its second argument yields
\begin{equation}
g(t,L,u) = \Phi(tL^{\frac{1}{\nu}}) + uL^{-\delta} \Phi_u(tL^{\frac{1}{\nu}})~.
\end{equation}
In the absence of corrections to scaling ($u=0$), the Binder cumulants at $t=0$ corresponding to different system sizes have
the universal value $\Phi(0)$, i.e., the critical temperature is marked by a crossing of all Binder cumulant curves.
If corrections to scaling cannot be neglected ($u \ne 0$), this is not the case (see, e.g., Ref.\ \cite{SelkeShchur05})
because $g(0,L,u)$ is not independent of $L$ but takes the value $g(0,L,u) = \Phi(0) + uL^{-\delta}\Phi_u(0)$.
Instead, the crossing point shifts with $L$ and approaches $t=0$ as $L \rightarrow \infty$.
The functional form of this shift can be worked out explicitly by expanding the scaling
functions $\Phi$ and $\Phi_u$,
\begin{equation}
g(t,L,u) = \Phi(0) + tL^{\frac{1}{\nu}}\Phi'(0) + uL^{-\delta}\Phi_u(0)~.
\end{equation}
Using this expression to evaluate the crossing temperature $T^*(L)$ between the Binder cumulant curves for sizes
$L$ and $cL$ (where $c$ is a constant) yields
\begin{equation}
T^*(L) = T_c+ b L^{-\omega} \quad \textrm{with} \qquad \omega = \delta+\frac{1}{\nu}
\label{eq:crossingT}
\end{equation}
where $b \sim u$ is a non-universal amplitude.

To determine the crossing temperature, we fit the $g$ vs $T$ data sets corresponding to different system
sizes with separate quartic polynomials.(Quartic polynomials provide reasonable fits within the temperature range
of interest while avoiding spurious oscillations.) The intersection point of these polynomials yields
the crossing temperature $T^*$. To estimate the errors of the crossing temperature we use an ensemble method.
For each $g(T)$ curve,
we create an ensemble of artificial data sets $g_a(T)$ by adding noise to the data
\begin{equation}
g_a(T) = g(T) + \Delta g(T)\, r ~.
\end{equation}
Here, $r$ is a random number chosen from a normal distribution of zero mean and unit variance, and $\Delta g(T)$ is
the statistical error of the Monte Carlo data for $g(T)$. Note that we use the same random number $r$ for the  entire $g(T)$ curve, leading to an upward or downward shift of the curve. This stems from the fact that the
statistical error $\Delta g(T)$ is dominated by the disorder
noise while the Monte Carlo noise is much weaker. This implies that the deviations at different temperatures
of the Binder cumulant from the true average are correlated.
Repeating the crossing analysis with these ensembles of curves, we get ensembles of crossing temperatures.
Their mean and standard deviation yield $T^*$ and the associated error $\Delta T^*$, respectively.

%%%%%%%%%%%%%%%%%%%%%%%%%%%%%%%%%%%%%%%%%%%%%%%%%%%%%%%%%%%%%%%%%%%%%%%%%%%%%%%%%%%%%%%%%%%%%%%%%%%%%%%%%%%%%%%%%
\section{Results}
%%%%%%%%%%%%%%%%%%%%%%%%%%%%%%%%%%%%%%%%%%%%%%%%%%%%%%%%%%%%%%%%%%%%%%%%%%%%%%%%%%%%%%%%%%%%%%%%%%%%%%%%%%%%%%%%%
In this section we report the results of our simulations for cubic and hexaferrite lattices occupied by
XY or Heisenberg spins.

\subsection{Cubic Lattices}
\label{sec:cubic}
%%%%%%%%%%%%%%%%%%%%%%%%%%%%%%%%%%%%%%%%%%%%%%%%%%%%%%%%%
%XY SPINS   %%%%%%%%%%%%%%%%%%%%%%%%%%%%%%%%%%%%%%%%%%%%%

We investigate the behavior of both XY and Heisenberg models on cubic lattices.
To check the validity of our simulations, we first consider clean (undiluted) lattices.
We find critical temperatures of $T_c = 2.2017(1)$ and $T_c = 1.44298(2)$ for XY and Heisenberg spins, respectively.
They agree well with previously known numerical results \cite{Gottlob93,Brown2006}.

\begin{figure}
\includegraphics[width=\linewidth]{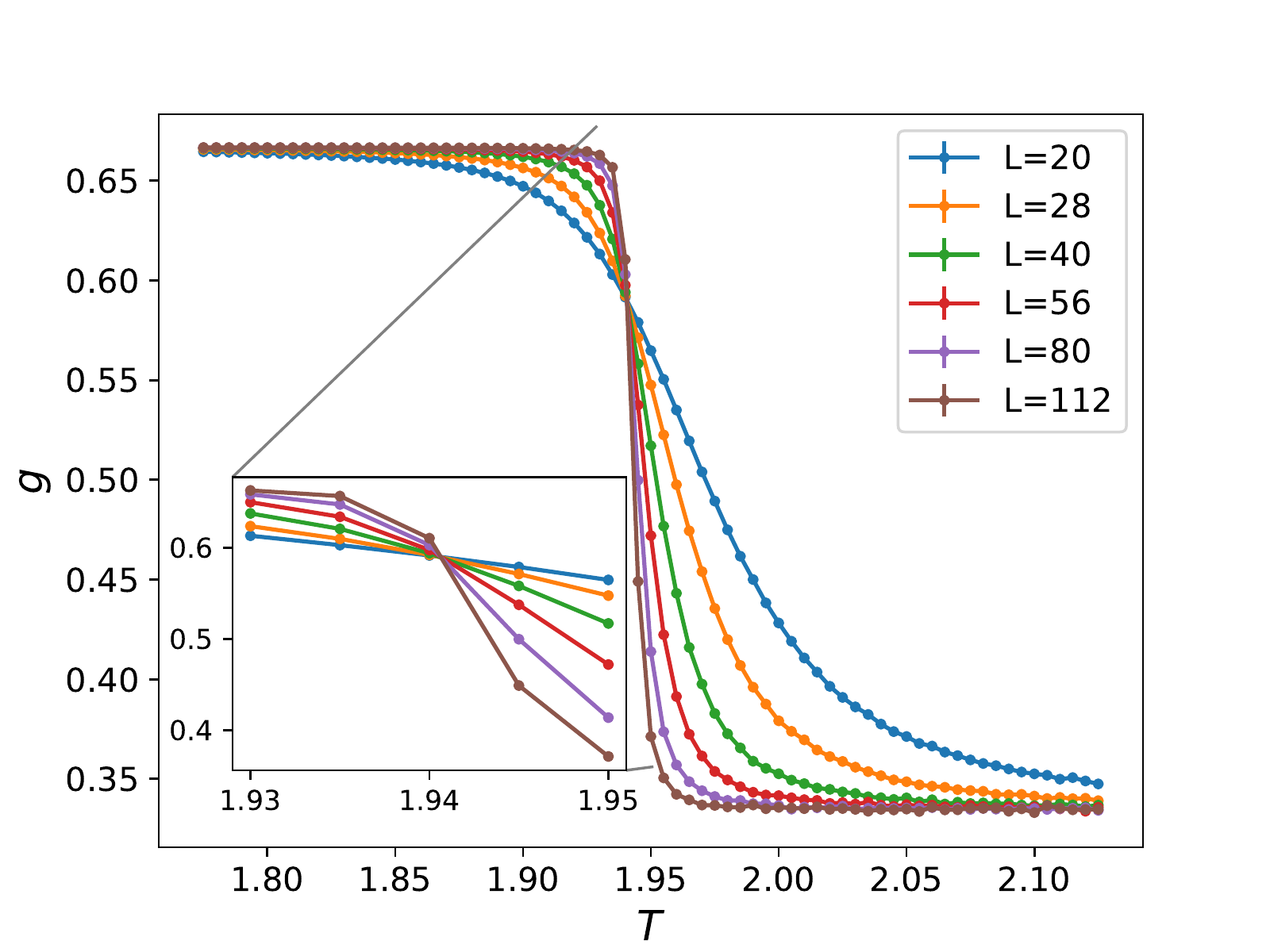}
\caption{Binder cumulant $g$ vs temperature $T$ for the cubic lattice XY model with dilution $p=0.10$.
The statistical errors arising from the Monte Carlo simulation are smaller than the symbol size.
The inset show the intersection region of the curves more closely. All curves cross at the same point
within their statistical errors.
}
\label{fig:BinderXY10}
\end{figure}

We now turn to diluted systems, starting with the XY case. For reference, the site
percolation threshold of the simple cubic lattice is at the vacancy probability $p_c=0.6883923(2)$ \cite{Wang13}.
For low dilutions ($p<0.64$), the Binder cumulant vs.\ temperature curves for all simulated system sizes
cross through exactly the same point within their statistical errors, implying that corrections to the leading
finite-size scaling behavior are not important. Therefore, we determine $T_c$ from the crossing of
the $g(T)$ curves of the two largest system sizes,
$L^3=80^3$ and $L^3=112^3$. The ensemble method is applied to find the error of $T_c$.
Fig. \ref{fig:BinderXY10} shows an example of this situation for dilution $p=0.1$.

\begin{figure}
\includegraphics[width=\linewidth]{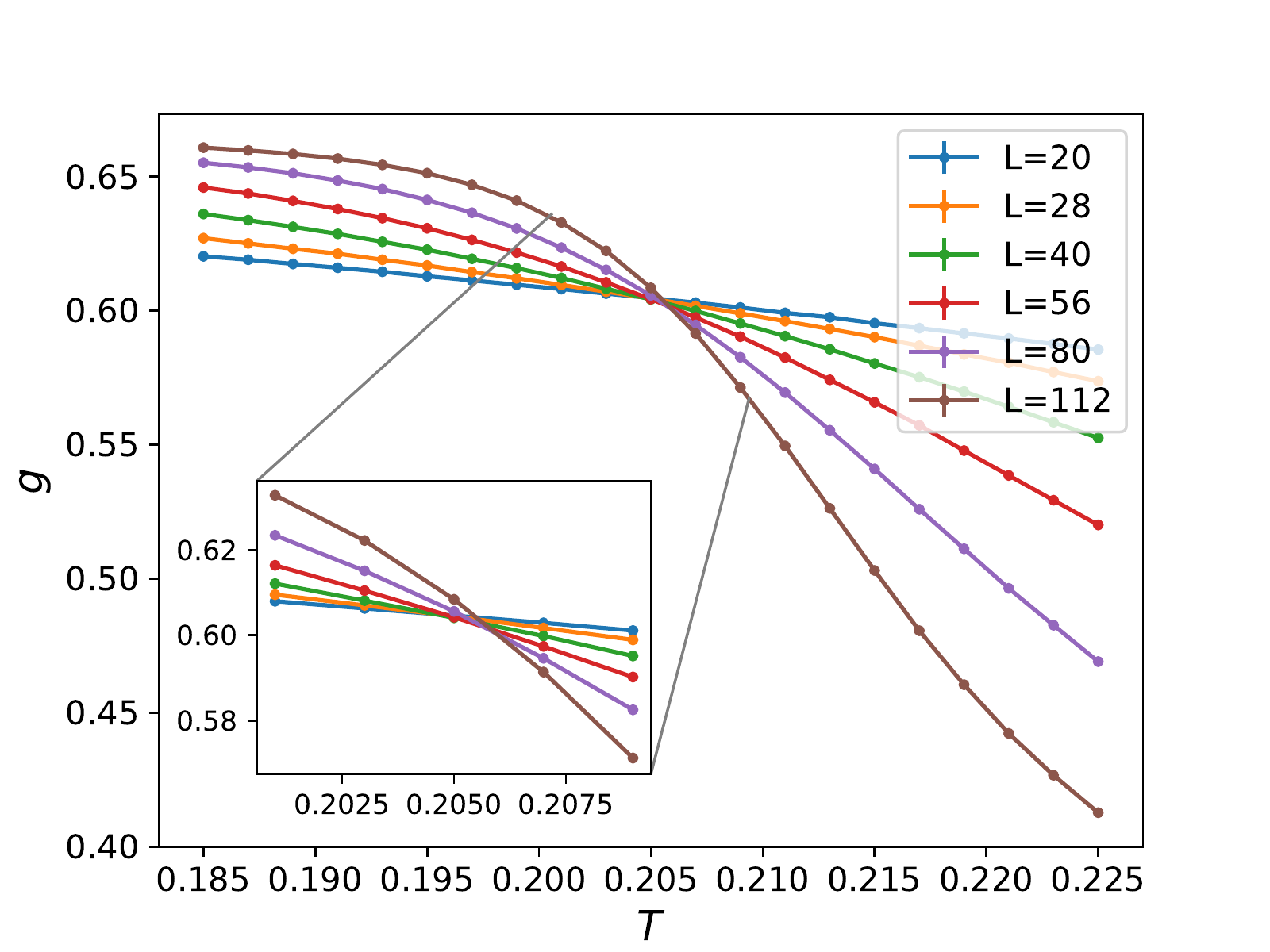}
\caption{Binder cumulant $g$ vs temperature $T$ for the XY model on a cubic lattice for dilution $p=0.65$, i.e. close to $p_c$.
The curves do not all cross at the same temperature. Instead, the crossing progressively shifts as $L$
increases. The statistical errors arising from the Monte Carlo simulation are smaller than the symbol size.}
\label{fig:BinderXY65}
\end{figure}

\begin{figure}
\includegraphics[width=\linewidth]{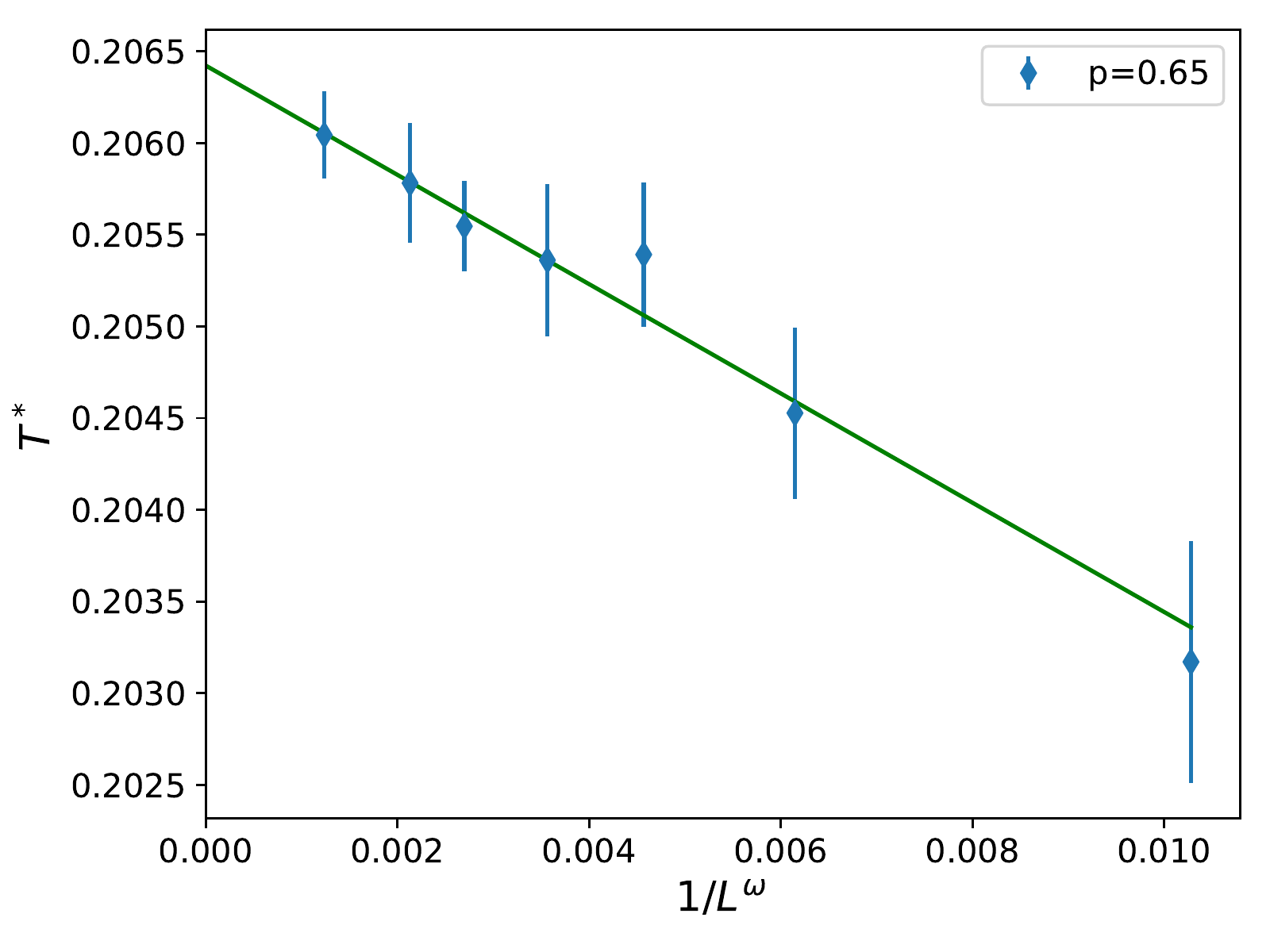}
\caption{Extrapolation to infinite system size of the crossing temperature $T^*$ of the Binder cumulant curves for
system sizes $L$ and $\sqrt{2}L$ using $\omega=1.5$. The dilution is $p=0.65$. A fit to Eq. \eqref{eq:crossingT} gives
$T_c =0.2064(4)$. The error bars of $T^*$ have been determined using the ensemble method described in Sec.\ \ref{subsec:data_analysis}.}
\label{fig:cfit}
\end{figure}

\begin{figure}
\includegraphics[width=\linewidth]{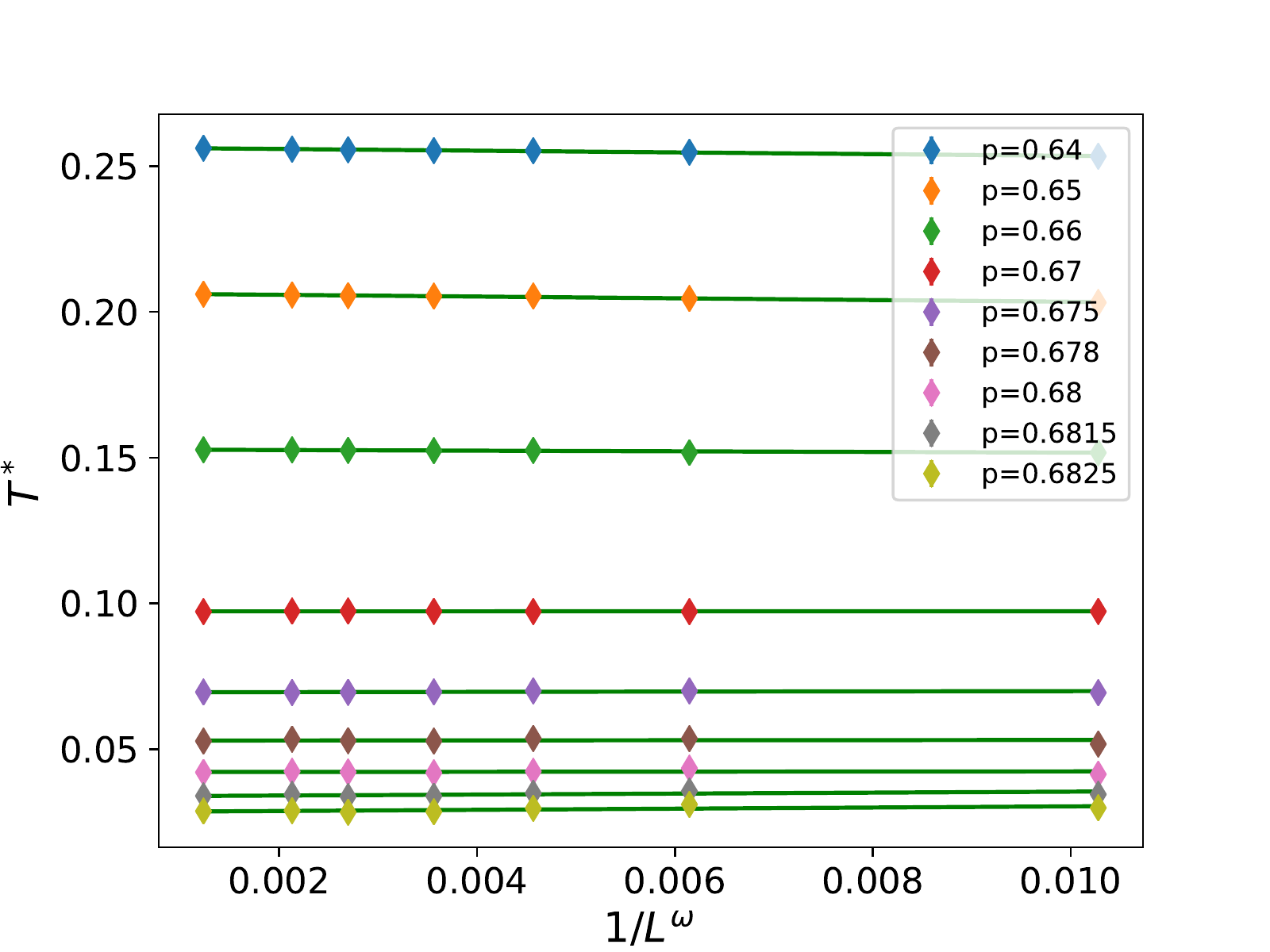}
\caption{Overview of the extrapolations of the crossing temperatures $T^*$ for several dilutions near $p_c$
using $\omega=1.5$. The error bars $\Delta T^*$ are smaller than the symbols.}
\label{fig:allcfit}
\end{figure}

For higher dilutions ($p \geq 0.64)$ in the vicinity of the percolation threshold $p_c$, the crossing of
the Binder cumulant vs.\ temperature curves is less sharp. Specifically, the crossing temperature $T^*(L)$
of the curves for linear system sizes $L$ and $\sqrt{2} L$ shifts visibly towards higher temperatures as
the system sizes are increased. An example (for $p=0.65$) is demonstrated in Fig. \ref{fig:BinderXY65}.
As shown in the previous section, this shift is caused by corrections to the leading
finite-size scaling behavior. According to Eq.\ (\ref{eq:crossingT}), it can be modeled as $T^*(L)=T_c+bL^{-\omega}$.
To find the asymptotic (infinite system size) value of $T_c$, we thus fit the crossing temperature $T^*(L)$ to Eq. (\ref{eq:crossingT}).
As $\omega$ is expected to be universal, i.e., to take the same value for all dilutions near $p_c$, we perform a combined fit for all dilutions $p \ge 0.64$
and treat $\omega$ as a fitting parameter. This combined fit produces $\omega=1.5 \pm 0.4$.
An example of the resulting extrapolation is presented in Fig. \ref{fig:cfit} for $p=0.65$.
The figure shows that the finite-size shifts of the crossing temperature are not very strong.
This is further confirmed in Fig.\ \ref{fig:allcfit} which presents an overview of the fits for
all dilutions from $p=0.64$ to $p=0.6825$.

\begin{figure}
\includegraphics[width=\linewidth]{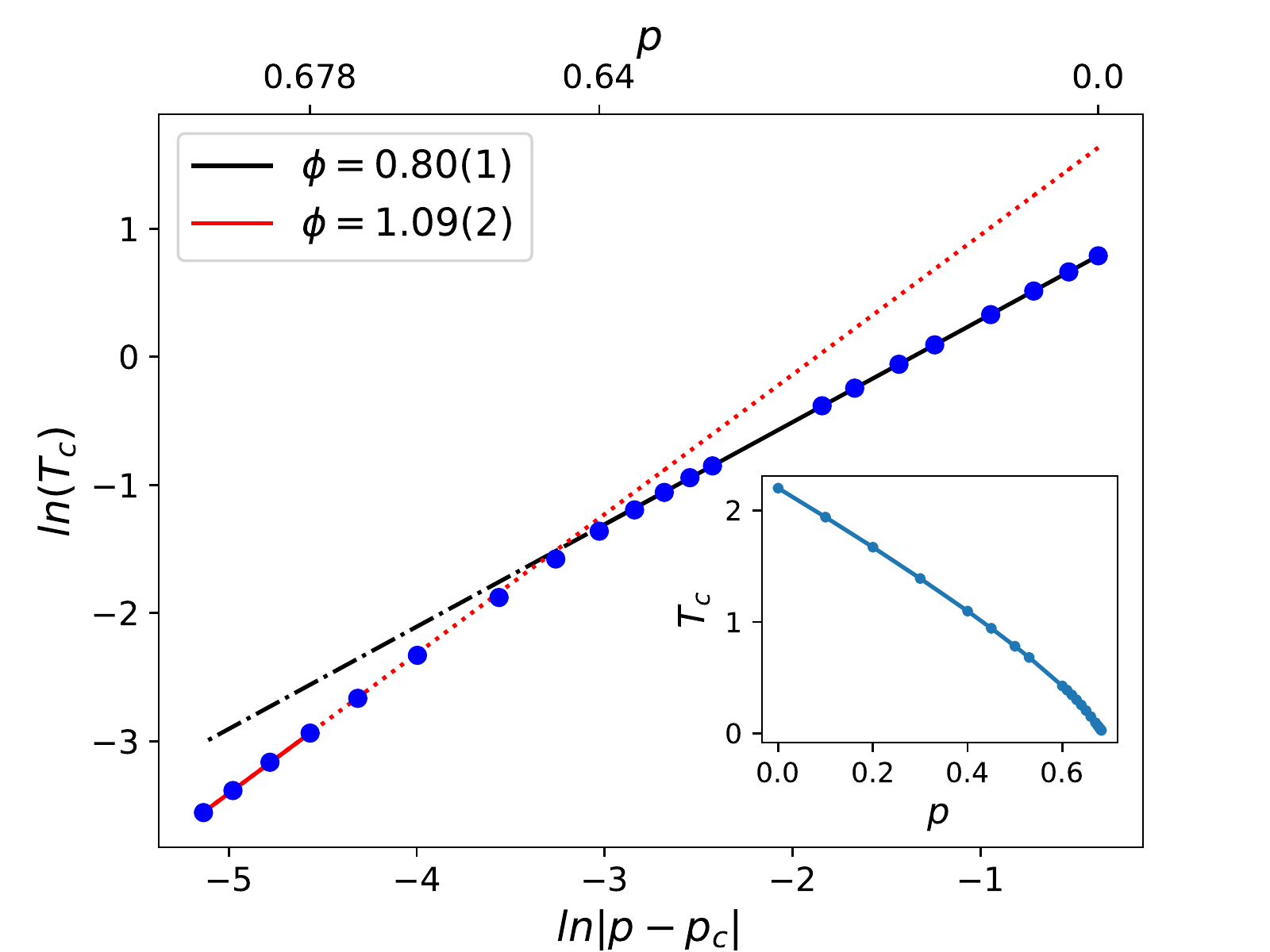}
\caption{Phase boundary of the site-diluted XY model on a cubic lattice. Main panel: Log-log plot of $T_c$ vs.\
$|p-p_c|$. The straight lines are power-law fits, $T_c \sim |p-p_c|^\phi$.  They are shown as solid lines within the fit range. The dotted and dash-dotted lines are extrapolations. For details see text. Inset: Overview presented as linear plot of $T_c$ vs.\ $p$. All error bars of the data points are
smaller than the symbol size.}
\label{fig:powerXY}
\end{figure}

The resulting phase boundary $T_c(p)$ of the site-diluted XY model on a cubic lattice is shown in
Fig.\ \ref{fig:powerXY}.
The overview given in the inset demonstrates that $T_c(p)$ is indeed continuously suppressed with increasing $p$ and approaches zero as $p \rightarrow p_c$.
To analyze the functional form of $T_c(p)$ close to $p_c$, the main panel of Fig. \ref{fig:powerXY} shows a log-log plot
of $T_c$ vs.\ $|p-p_c|$. We observe that the phase boundary follows two different power laws, close to the percolation threshold $p_c$ and further away from $p_c$. The asymptotic value of $\phi$ is determined from a fit of the data closest to $p_c$ (viz.\ $p$ between $0.678$ to $0.6825$), yielding a crossover exponent of $\phi=1.09(2)$. Its error estimate is a combination of the statistical error from the fit and a systematic error estimated from the robustness of the value against changes of the fit interval. The asymptotic value of $\phi$ agrees reasonably well with the prediction of percolation theory.
The asymptotic power law describes the data for dilutions above about $p=0.65$. The asymptotic critical region thus ranges
from about $p=0.65$ to $p_c=0.6883923$.

The preasymptotic behavior of $T_c(p)$ for $p$ between $p=0$ to $p=0.64$ also follows a power law in good approximation.
However, the exponent is significantly below unity, $\phi = 0.80(1)$.

%%%%%%%%%%%%%%%%%%%%%%%%%%%%%%%%%%%%%%%%%%%%%%%%%%%%%%%%
%HEISENBERG SPINS%%%%%%%%%%%%%%%%%%%%%%%%%%%%%%%%%%%%%%%%

\begin{figure}
\includegraphics[width=\linewidth]{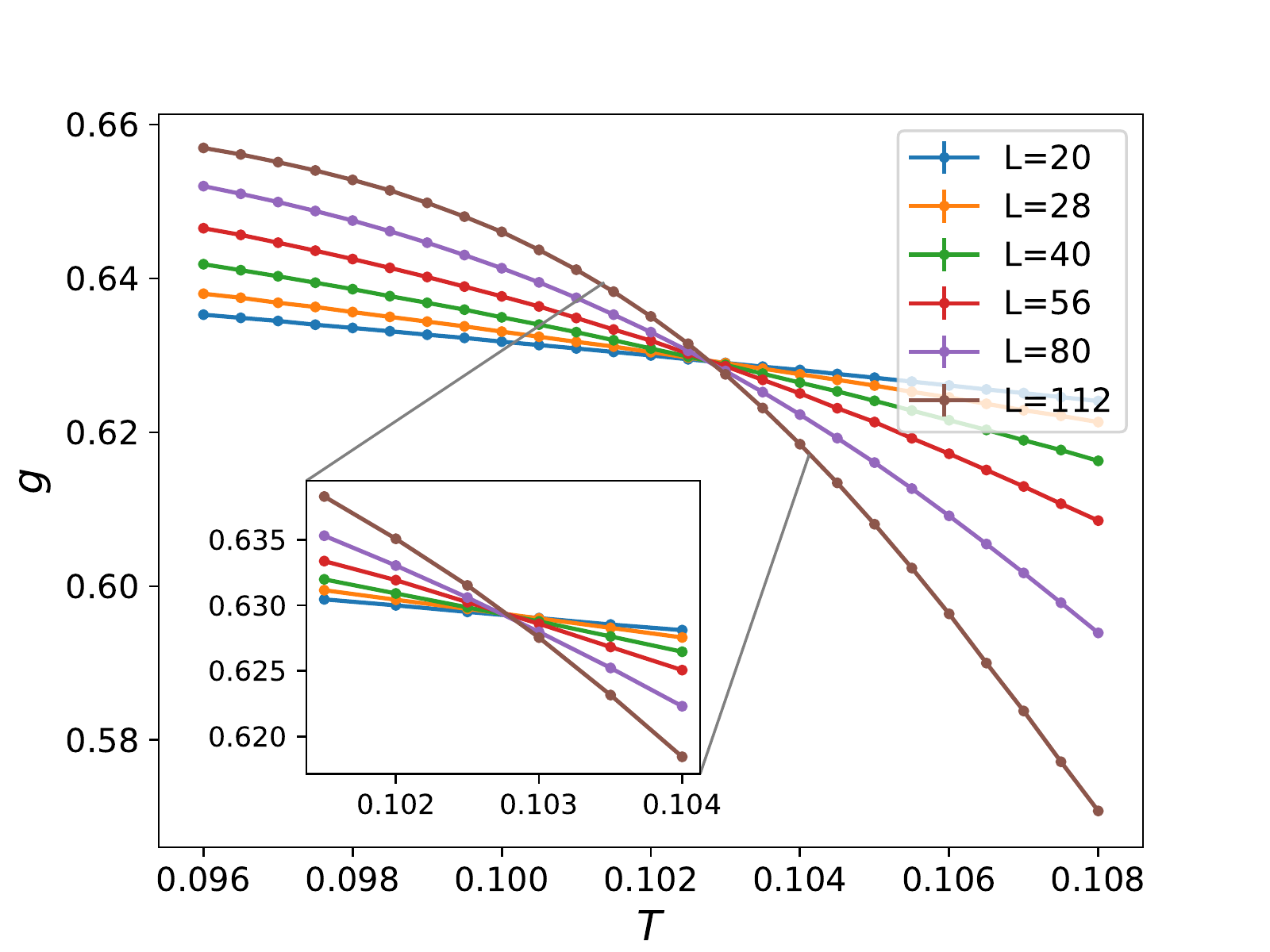}
\caption{Binder cumulant $g$ vs temperature $T$ for dilution $p=0.65$ on cubic lattice and Heisenberg spins. All curves
cross at the same temperature. Error bars are smaller than the symbol size.}
\label{fig:BinderH65}
\end{figure}

We proceed in the same manner for the Heisenberg model on the cubic lattice. Starting from the clean case, we gradually
increase dilution and find $T_c(p)$. In the case of Heisenberg spins, we find that the corrections to finite-size scaling
are weaker than in the XY case.
Even in the vicinity of $p_c$, all Binder cumulant curves intersect in a single point within their statistical errors.
As an example, the $g$ vs $T$ data for $p=0.65$ are shown in Fig.\ \ref{fig:BinderH65}.
The critical temperatures $T_c(p)$ and its error are therefore determined from the Binder cumulant crossing for system
sizes $L^3=80^3$ and $L^3=112^3$, the largest systems simulated.

\begin{figure}
\includegraphics[width=\linewidth]{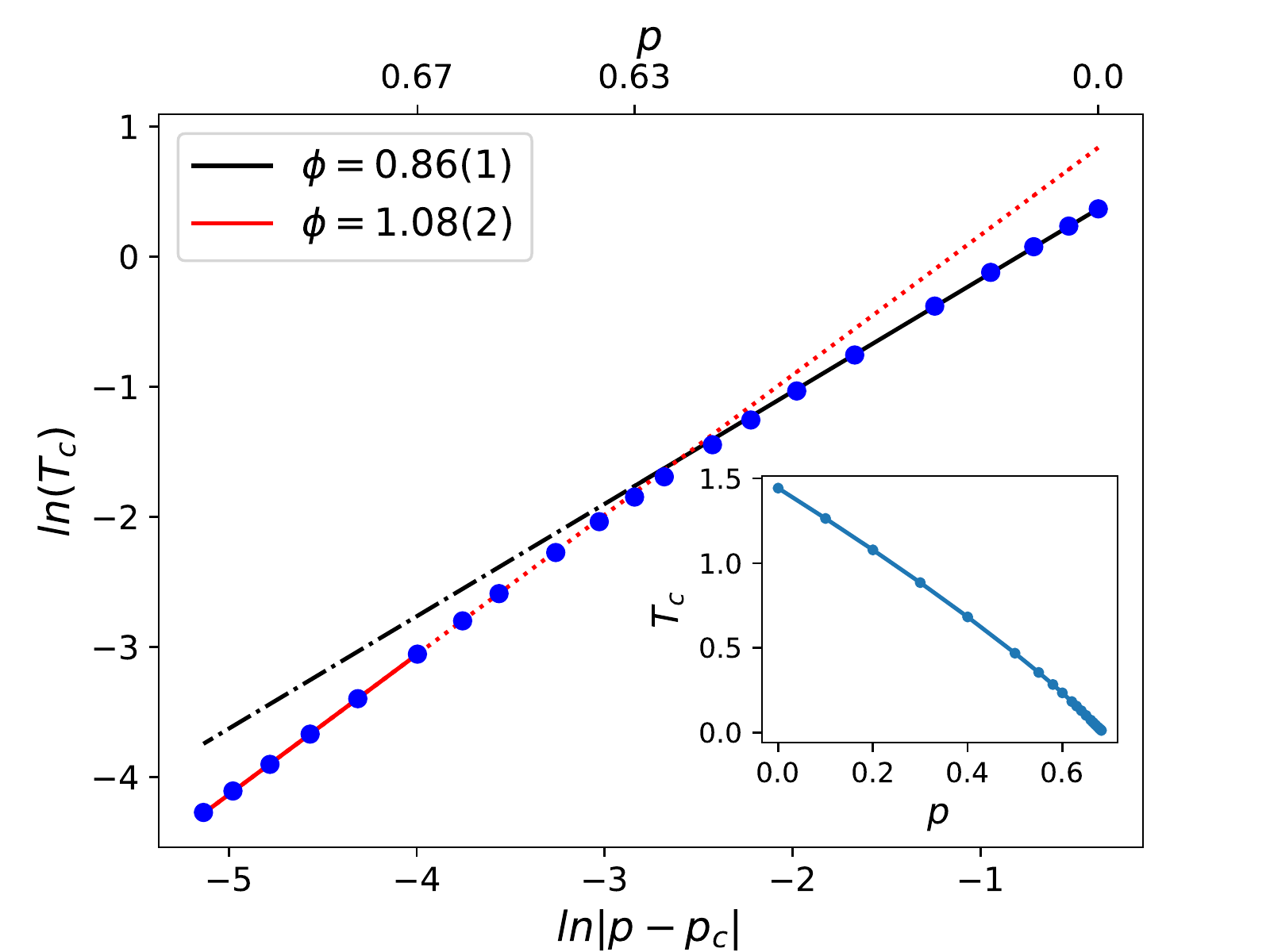}
\caption{Phase boundary of the site-diluted Heisenberg model on a cubic lattice. Main panel: Log-log plot of $T_c$ vs.\
$|p-p_c|$. The straight lines are power-law fits, $T_c \sim |p-p_c|^\phi$.  They are shown as solid lines within the fit range. The dotted and dash-dotted lines are extrapolations. For details see text. Inset: Overview presented as linear plot of $T_c$ vs.\ $p$. All error bars of the data points are
smaller than the symbol sizes.}\label{fig:powerH}
\end{figure}

The phase boundary of the site-diluted Heisenberg model on a cubic lattice is constructed from these data and shown in
Fig. \ref{fig:powerH}.
Similar to the XY case, we observe two separate power law exponents governing the phase boundary. The
dilutions $p \gtrsim 0.65$ constitute the asymptotic critical region with crossover exponent $\phi=1.08(2)$, in agreement
with the percolation theory prediction. The nonuniversal preasymptotic crossover exponent obtained for dilutions
$p \lesssim 0.62$ is again smaller than unity, $\phi=0.86(1)$, but somewhat larger than in the XY case.

%%%%%%%%%%%%%%%%%%%%%%%%%%%%%%%%%%%%%%%%%%%%%%%%%%%%%%%%%%%%%%%%%%%%%%
\subsection{Hexagonal Ferrite Lattice}
\label{sec:hexaferrite}
%%%%%%%%%%%%%%%%%%%%%%%%%%%%%%%%%%%%%%%%%%%%%%%%%%%%%%%%%%%%%%%%%%%%%%

Whereas the asymptotic critical behavior of the phase boundary close to the percolation threshold
is expected to be universal, its behavior outside the asymptotic critical region does not have to be universal.
For a better quantitative understanding of the magnetic phase boundary of the diluted hexaferrites, we therefore
also perform simulations of the Heisenberg model (\ref{eq:H_hexa}) using the hexaferrite crystal structure
and realistic exchange
interactions. In the calculations, we focus on the leading non-frustrated interactions, as outlined
in Sec.\ \ref{subssec:model-hexa}. The site percolation threshold for the lattice spanned by these
interactions is $p_c = 0.7372(5)$ \cite{Rowleyetal17}.

\begin{figure}
\includegraphics[width=\linewidth]{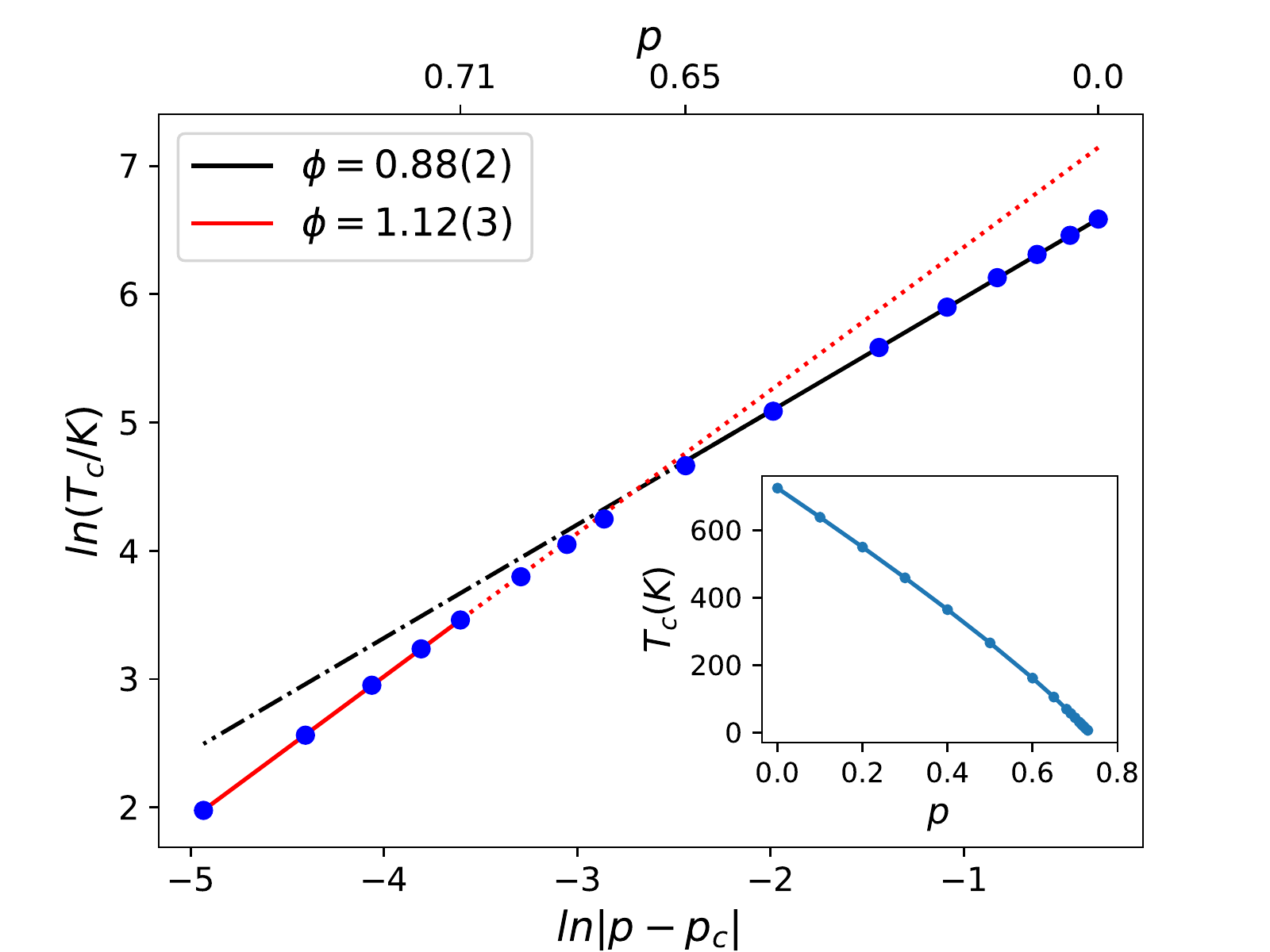}
\caption{Phase boundary for the Heisenberg model on a hexagonal ferrite lattice. The main panel shows the log-log plot of $T_c$ vs.\ $|p-p_c|$. The statistical errors of the data (determined by the ensemble method) are smaller than the symbol size. The straight lines are fits to $T_c \sim |p-p_c|^\phi$.  They are shown as solid lines within the fit range. The dotted and dash-dotted lines are extrapolations. For details see text.
The inset shows a linear plot the complete phase boundary $T_c(p)$.}
\label{fig:powerhexa}
\end{figure}

As before, the critical temperature $T_c$ for a given dilution is determined from the Binder cumulant crossings.
Corrections to the finite-size scaling were found to be negligible within the statistical errors.
Thus, we used the Binder cumulant crossing of the two largest system sizes ($28^3$  and $40^3$ double unit cells)
to find $T_c$. The resulting phase boundary is shown in Fig.\ \ref{fig:powerhexa}.
The behavior of this phase boundary is very similar to the cubic lattice results.
High dilutions, $p \gtrsim 0.68$, fall into the asymptotic critical region with a crossover exponent of $\phi=1.12(3)$,
in excellent agreement with the percolation theory predictions. This also confirms the universality of the asymptotic crossover exponent.  The preasymptotic exponent $\phi=0.88(2)$ that governs the behavior for dilutions below about
0.65 is smaller than unity and takes roughly the same value as for the Heisenberg model on the cubic lattice.

\begin{figure}
\includegraphics[width=\linewidth]{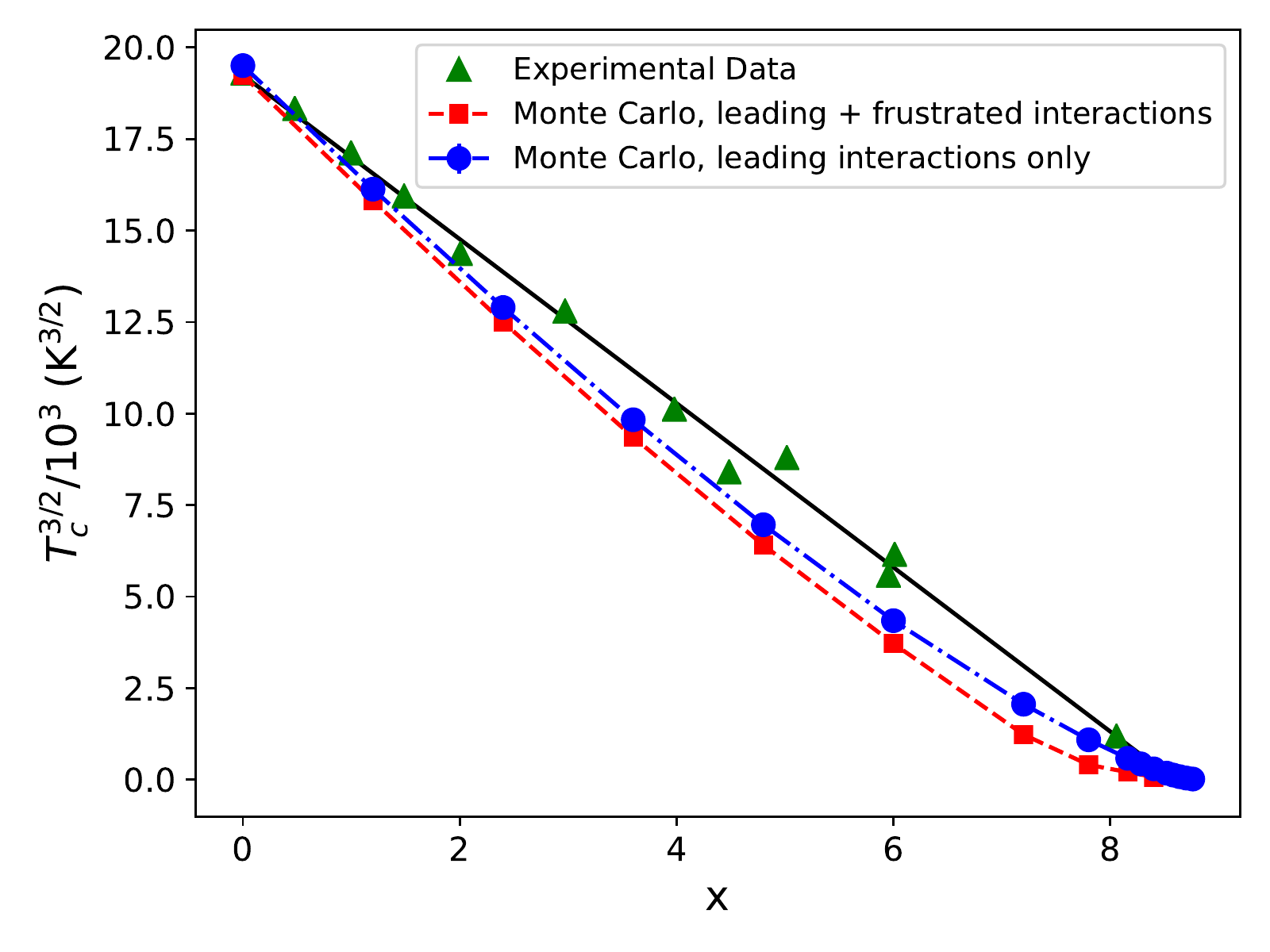}
\caption{Comparison between the numerically determined phase boundary $T_c(x)$ and the experimental data
for PbFe$_{12-x}$Ga$_x$O$_{19}$ \cite{Rowleyetal17}.
The tuning parameter $x$ is related to the dilution by $x/12 = p$.
The Monte Carlo simulations show a more rapid suppression of $T_c$ with $x$. Including additional weak
frustrated interactions increases the discrepancy.}
\label{fig:phaseBcompare}
\end{figure}

Our numerical results disagree with the experimentally observed 2/3 power law, $T_c (x) = T_c(0) (1-x/x_c)^{2/3}$.
In the simulations, the transition temperature $T_c$ is suppressed more rapidly with $x$ than in the
experimental data (see Fig.\ \ref{fig:phaseBcompare}).
To explore possible reasons for this discrepancy, we also perform test simulations that include additional weaker
exchange interactions \cite{Wuetal16} that frustrate the ferrimagnetic
order. The results of these simulations, which are included in Fig.\ \ref{fig:phaseBcompare}, show that these weaker frustrating interactions have little effect at low dilutions.
At higher dilutions, when the ferrimagnetic order is already weakened, the frustrating interactions further suppress
the transition temperature. They thus further increase the discrepancy between the experimental data
and the Monte Carlo results.

%%%%%%%%%%%%%%%%%%%%%%%%%%%%%%%%%%%%%%%%%%%%%%%%%%%%%%%%%%%%%%%%%%%%%%%%%%%%%%%%%%%%%%%%%%%%%%%%%%%%%%%%%%%%%%%%
\section{Conclusion}
\label{sec:conclusions}
To summarize, motivated by recent experimental observations on hexagonal ferrites,
we have studied classical site-diluted XY
and Heisenberg models by means of large-scale Monte Carlo simulations, focusing on the shape of the magnetic phase
boundary. We have obtained two main results.

First, for high dilutions close to the lattice percolation threshold, the critical temperature depends on the
dilution via the power law $T_c \sim |p-p_c|^{\phi}$ in all studied systems. In this asymptotic region, we have found the values $\phi = 1.09(2)$ and 1.08(2) for XY and Heisenberg spins on cubic lattices, respectively. For the Heisenberg model
on the hexaferrite lattice, $\phi = 1.12(3)$. These values agree with each other and with the prediction $\phi=1.12(2)$ of classical percolation theory. The crossover exponent $\phi$ thus appears to be super-universal, i.e., it takes the same
value not just for different lattices but also for XY and Heisenberg symmetry.

Interestingly, the asymptotic critical region of the percolation transition
is very narrow, as the asymptotic power-laws only hold in the range
$|p-p_c| \lesssim 0.04$. At lower dilutions, the phase boundary still follows a power law in $|p-p_c|$, but with an exponent that appears to be non-universal and below unity (in the range between 0.8 and 0.9).

Our second main result concerns the origin of the $2/3$ power law,
$T_c (x) = T_c(0) (1-x/x_c)^{2/3}$, that was experimentally observed in PbFe$_{12-x}$Ga$_x$O$_{19}$
over the entire concentration range between 0 and close to the percolation threshold \cite{Rowleyetal17}.
Neither the asymptotic nor the preasymptotic power laws identified in the simulations match the experimental
result. In fact, in all simulations, the critical temperature is suppressed more rapidly with increasing dilution
than in the experiment. The observed shape of the magnetic phase boundary in PbFe$_{12-x}$Ga$_x$O$_{19}$
thus remains unexplained.

Potential reasons for the unusual behavior may include the interplay between magnetism and ferroelectricity in these materials \cite{Rowleyetal16}
or the presence of quantum fluctuations (arising from the frustrated magnetic
interactions mentioned above), even though it is hard to imagine that these stay relevant at temperatures as high as 720\,K.
Another possible explanation could be a statistically unequal occupation of the different iron sites in the unit cell by Ga ions.
Exploring these possibilities
remains a task for the future. Disentangling these effects may also require additional experiments introducing
further tuning parameters such as pressure or magnetic field in addition to chemical composition.

%%%%%%%%%%%%%%%%%%%%%%%%%%%%%%%%%%%%%%%%%%%%%%%%%%%%%%%%%%%%%%%%%%%%%%%%%%%%%%%%%%%%%%%%%%%%%%%%%%%%%%%%%%%%%%%%
\begin{acknowledgements}
We acknowledge support from the NSF under Grant Nos. DMR-1506152, DMR-1828489, and OAC-1919789.
The simulations were performed on the Pegasus and Foundry clusters at Missouri S\&T.
We also thank Martin Puschmann for helpful discussions.
\end{acknowledgements}

\def\bibfont{\footnotesize}

%\bibliographystyle{spphys-tv}
\bibliographystyle{apsrev4-1}
%\bibliography{../00bibtex/rareregions}   % name your BibTeX data base
\bibliography{percolation}   % name your BibTeX data base

\end{document}